\documentclass{ws-ijmpb}

\begin{document}

\markboth{Ettouhami, Klironomos and Dorsey}
{Possible new bubble phases of 2D electrons in higher Landau levels}

\catchline{}{}{}{}{}

\title{Possible new bubble phases of two-dimensional electrons \\
in higher Landau levels}

\author{\underline{A.M. Ettouhami}\footnote{Email: mouneim@phys.ufl.edu},
F.D. Klironomos and A.T. Dorsey}
\address{Department of Physics, University of Florida, P.O. Box 118440, Gainesville, FL 32611}


\maketitle

\begin{history}
\end{history}

\begin{abstract}

We argue that, due to the specific form of the interaction potential between electronic guiding 
centers, a bubble crystal (BC) with basis may be energetically more favorable than the usual 
bubble solid. This new BC has well-defined normal eigenmodes, and remains energetically favorable 
when the effects of finite sample thickness and of screening by electrons in lower LLs are taken 
into account.

\end{abstract}

\keywords{Two-dimensional electron gas, Wigner crystal, Bubble states}


\section{Introduction}

Quantum phases of the two-dimensional electron gas in higher Landau levels 
have been the subject of intense study in recent years. Hartree-Fock
studies\cite{Fogler} have shown that, for small partial filling factors 
$\nu^* = \nu - 2n\approx 0.1-0.2$ (with $\nu$ the total filling 
factor and $n$ the Landau level index), the electrons form a triangular 
Wigner crystal, while for $\nu^*$ close to $1/2$ the ground state of the system is a 
stripe state. Between these two regions, a new crystalline solid
with more than one electron per lattice site, termed bubble crystal, 
turns out to be energetically more favorable. In this paper, we investigate
the interaction of electrons within individual bubbles, and argue that 
a bubble lattice with basis may lead to energetically more favorable ground 
states of the two-dimensional electron system, 
the properties of which we shall briefly describe.

\section{Wigner crystal and bubble phases in higher Landau levels}

We shall start from the expression of the Hartree-Fock energy 
of the partially filled $n$th Landau level, which is 
given by: 
\begin{eqnarray}
E_{HF} = \frac{1}{2}\int \frac{d\bf q}{(2\pi)^2}\;
V_{HF}({\bf q})\,|\rho({\bf q})|^2 \,,
\label{EHF}
\end{eqnarray}
where $V_{HF}$ is the Hartree-Fock
interaction, and the order parameter $\rho({\bf q})$ is 
the Fourier transform of the guiding center density, and is given by
$
\rho({\bf q}) = n({\bf q})/[e^{-q^2\ell^2/4}L_n(q^2\ell^2/2)]\,.
$
In this last equation, $n({\bf q})$ is the Fourier transform of the electron density, which 
we shall approximate by
$n({\bf r}) = \sum_{i,m} |\varphi_{nm}({\bf r} - {\bf R}_i)|^2$,
where $\varphi_{nm}({\bf r})$ is the noninteracting wavefunction of angular momentum $m$
and Landau level index $n$, and where the summation extends over all the bubbles located at the 
lattice sites ${\bf R}_i$ of a triangular Bravais lattice and all the electrons within each bubble. 
The Hartree-Fock interaction potential in Eq. (\ref{EHF})
consists of the sum of a Hartree and Fock parts, which are given by:
\begin{eqnarray}
V_H({\bf q}) & = & \frac{2\pi e^2}{\epsilon q}\;\mbox{e}^{-q^2\ell^2/2}
\Big[L_n(q^2\ell^2/2)\Big]^2 \,,
\label{VH}
\\
V_F({\bf q}) & = & - (2\pi\ell^2)\;\int\frac{d^2{\bf q}'}{(2\pi)^2} 
V_H({\bf q}')\,e^{-i{\bf q}\times{\bf q}'\ell^2} 
\label{VF}
\,.
\end{eqnarray}
In the above expressions, $\epsilon$ is the dielectric constant of
the host semiconductor, and $L_n(x)$ is the $n$th Laguerre polynomial.

It can easily be checked that the projected density $\rho({\bf q})$ can be written in the form
$
\rho({\bf q})=\sum_m\tilde{\rho}_m({\bf q}) \sum_i e^{-i{\bf q}\cdot{\bf R}_i}\,,
$
with the partial guiding center density
$\tilde{\rho}_m({\bf q})=\tilde{n}_m({\bf q})/[e^{-q^2\ell^2/4}L_n(q^2\ell^2/2)]$ (here
$\tilde{n}_m({\bf q})=\int d{\bf r}|\varphi_{nm}({\bf r})|^2e^{-i{\bf q}\cdot{\bf r}}$ 
is the Fourier transform of the density of electrons in state $m$). Using the above expression of 
$\rho({\bf q})$ into Eq. (\ref{EHF}) allows us to rewrite the cohesive energy $E_{HF}$
in the form:
\begin{eqnarray}
E_{HF} = \frac{1}{2}\,
\sum_{i\neq j}\sum_{m,m'} U_{m,m'}({\bf R}_i - {\bf R}_j) +
\sum_{i}\sum_{m < m'} U_{m,m'}(0) \,,
\label{newEHF}
\end{eqnarray}
where we introduced the interaction potential $U_{m,m'}$ between electrons
in states $m$ and $m'$, which in real space is given by:
\begin{eqnarray}
U_{m,m'}({\bf r}) = \int\frac{d^2\bf q}{(2\pi)^2}\,\rho_m({\bf q})V_{HF}({\bf q})\rho_{m'}({\bf q})
e^{i{\bf q}\cdot{\bf r}} \, .
\label{Ueff}
\end{eqnarray}
It can be verified that the above interaction potentials, when used in Eq. (\ref{newEHF}), 
give cohesive energies that are in excellent agreement with the Hartree-Fock calculations of 
Refs. 2 and 3.

\section{Lattice with basis: possible new bubble states}

\begin{figure}[ht]
\includegraphics[scale=0.7]{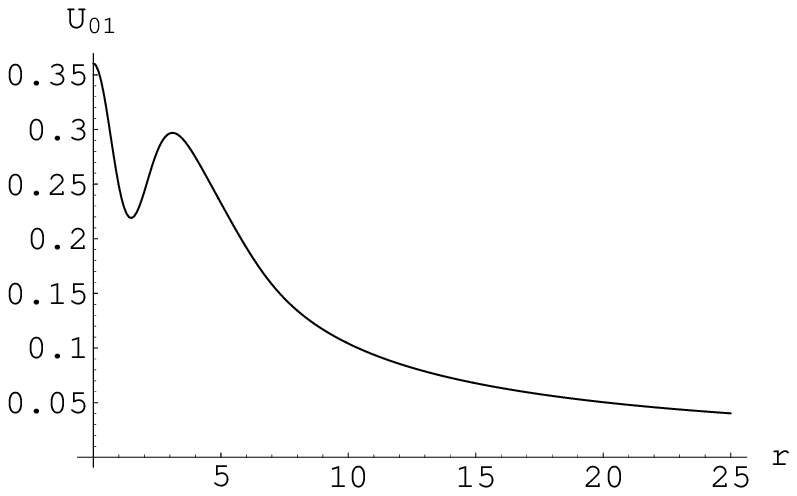}
\includegraphics[width=5.63cm,height=3.5cm]{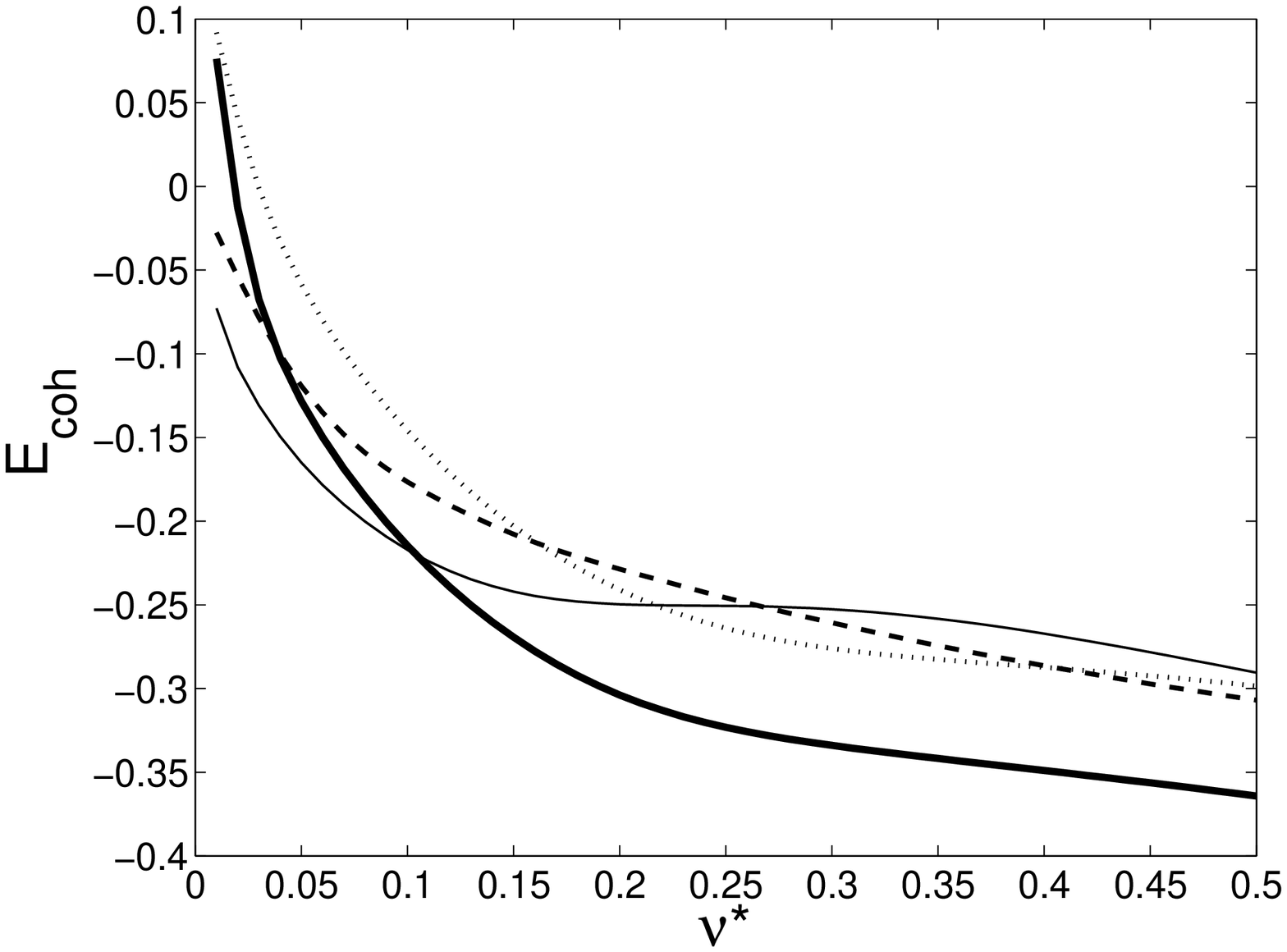}
\caption[]{
Left panel: Interaction potential $U_{01}$ (in units of $e^2/\epsilon\ell$) vs. $r$ (in units of 
$\ell$). Right panel: Cohesive energies (in units of $e^2/\epsilon\ell$)
of the Wigner crystal (thin solid line),
ordinary bubble crystal (dotted line), stipe phase (dashed line) and of the bubble crystal with a 
basis (thick line). The cohesive energy of the stripes is calculated as in Ref. 3.
Both figures are for Landau level $n=2$.
}\label{fig1}
\end{figure}

A plot of the interaction potential $U_{01}({\bf r})$ (through which the two electrons on a given
bubble interact) reveals the existence of a local minimum at $r=r_0=1.48\ell$
(see Fig.~\ref{fig1}),
which might favor a finite separation between the guiding centers of the two electrons in a 
given bubble. Aiming at finding an energetically more favorable ground state of the 
bubble type for our two-dimensional electron gas, the simplest trial ground state one can think 
of consists of a triangular lattice of bubbles with two electrons
per lattice site, but with the electrons guiding centers ${\bf r}_{mi}$ at lattice site 
${\bf R}_i$ separated from each other by a distance $r_0$, which may be
obtained by writing
${\bf r}_{mi}={\bf R}_i+(m-\frac{1}{2})r_0\hat{\bf x}$ (with $m=0,\,1$).
Such a ground state considerably lowers the cohesive energy of the two-electron bubble crystal,
as can be seen in Fig.~\ref{fig1}.

\section{Normal modes and stability of the bubble crystal}

\begin{figure}[t]
\includegraphics[width=5.63cm,height=3.5cm]{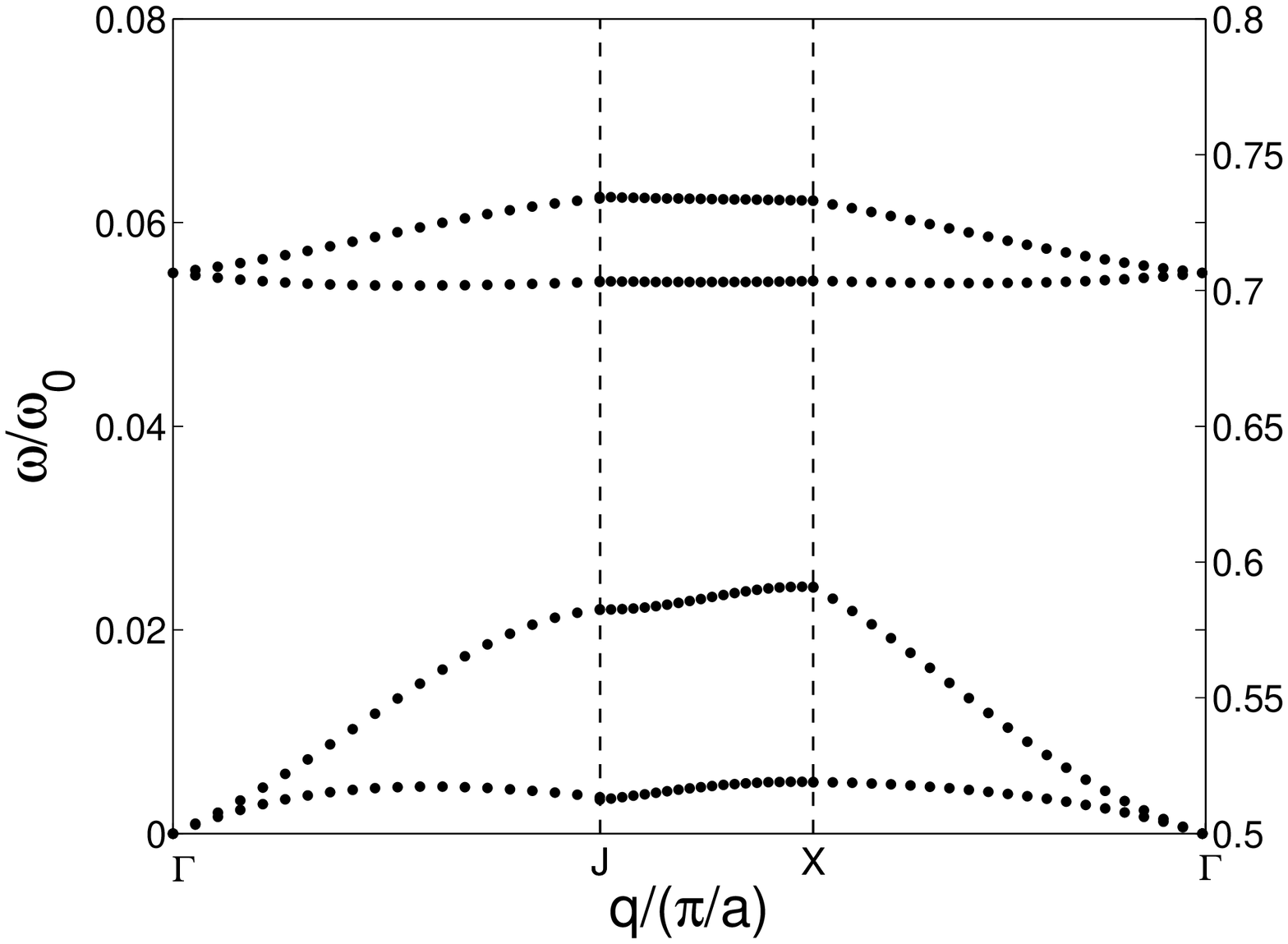}
\includegraphics[width=5.63cm,height=3.5cm]{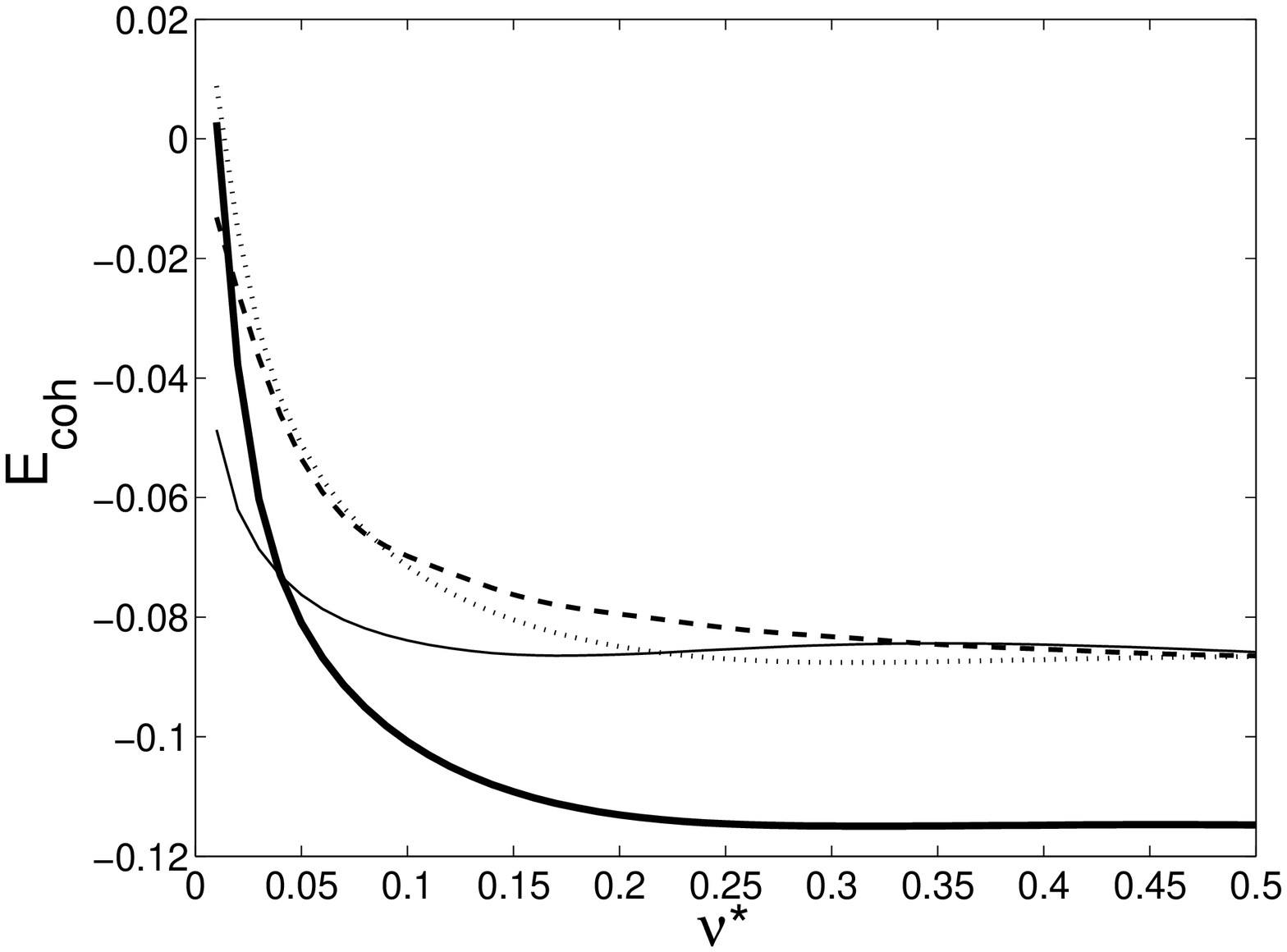}
\caption[]{
Left panel: Magnetophonon and magnetoplasmon dispersion curves along the boundary
of the irreducible element of the first Brillouin zone of the BC with basis
at $\nu=4.20$ ($a$ is the bubble lattice constant and $\omega_0=e^2/\hbar\epsilon\ell$).
The left hand side frequency scale is for magnetophonons, and the right one is for 
magnetoplasmons.
Right panel: same as the right panel in Fig. \ref{fig1}, but with
the $q$-dependent dielectric constant $\epsilon(q)$ and with the  finite sample 
thickness parameter $\lambda=1$. 
}\label{fignormalmodes}
\end{figure}

We now want to check the stability of our bubble crystal with basis. We shall do so by finding 
its normal vibration modes, and verifying that they are all well defined.
The derivation of these normal modes will proceed in a standard 
way as follows. To fix ideas, we shall consider the simplest case of the two-electron bubble 
crystal in the $n=2$ Landau level, and write 
for the electron guiding centers ${\bf r}_{mi}$ ($m=0,\,1$)  
at a given lattice site ${\bf R}_i$ the decomposition 
${\bf r}_{mi}={\bf R}_i+ (m-\frac{1}{2})r_0\hat{\bf x}+{\bf u}_{mi}$, 
where ${\bf u}_{mi}$ is the displacement of the $m$-th electron from the equilibrium lattice 
position ${\bf R}_{i}$. 
Expanding the cohesive energy $E_{HF}= \frac{1}{2}\,
\sum_{i\neq j}\sum_{m,m'} U_{m,m'}({\bf r}_{mi} - {\bf r}_{m'j}) +
\sum_{i}\sum_{m < m'} U_{m,m'}((m-m')r_0\hat{\bf x}+{\bf u}_{mi}-{\bf u}_{m'i})$ to second order 
in the small displacements ${\bf u}_{mi}$ leads to the elastic energy
$
E_{el} = \frac{1}{2}\sum_{i,j} \sum_{m,m'}
u_{m\alpha}({\bf R}_i)\Phi^{m,m'}_{\alpha\beta}({\bf R}_i-{\bf R}_j)u_{m'\beta}({\bf R}_j) \,,
$
where the elastic matrix $\Phi$ is given by
$
\Phi^{mm'}_{\alpha\beta}({\bf R}_i-{\bf R}_j) = \delta_{ij}\delta_{mm'}
\sum_{k,m''}\partial_{\alpha}\partial_{\beta}U_{mm''}({\bf R}_k +(m-m'')r_0\hat{\bf x}) 
-\partial_{\alpha}\partial_{\beta}U_{mm'}({\bf R}_i-{\bf R}_j+(m-m')r_0\hat{\bf x})\,.
$
In presence of a magnetic field, the 
equation of motion for the $m$-th electron at lattice site ${\bf R}_i$ has the form
$
m^*\frac{d^2}{dt^2}\,u_{m\alpha}({\bf R}_i) =
-\sum_{j,m'}\Phi_{\alpha\beta}^{m,m'}({\bf R}_i-{\bf R}_j)u_{m'\beta}({\bf R}_j)
-\frac{eB}{c}\varepsilon_{\alpha\beta}\frac{d}{dt}\,u_{m\beta}({\bf R}_i) \,
\label{eqmot}
$
($m^*$ is the effective mass of the electron in the host semiconductor 
and $\varepsilon_{\alpha\beta}$ is the complete antisymmetric 2D tensor).
We shall seek a solution to the above equation of motion 
that represents a wave with angular frequency
$\omega$ and wavevector ${\bf q}$,
$
u_{m\alpha}({\bf R}_i) = A_{m\alpha}({\bf q})\,e^{i({\bf q}\cdot{\bf R_i}-\omega t)} \;.
$
Substituting the above 
expression into the equation of motion, and solving the resulting secular equation leads to the 
normal eigenmodes plotted for $\nu=4.20$ in Fig. \ref{fignormalmodes} (left panel). 
As can be seen, the normal modes are all real, and consist as is well-known of two magnetophonon 
modes with characteristic dispersion $\omega(q\to 0)\sim q^{3/2}$, and two 
magnetoplasmon modes with $\omega(q\to 0)=\omega_c$ ($\omega_c=eB/m^*c$ being the cyclotron 
frequency).

We have also investigated the combined effects of screening by electrons from 
lower LLs (encoded in a wavevector-dependent dielectric constant $\varepsilon(q)$) and of a
finite sample thickness on the cohesive energy of the bubble crystal with basis. The 
\noindent
right panel on Fig. \ref{fignormalmodes} shows the cohesive energies of the
various phases vs. $\nu^*$ (at LL $n=2$), with the Coulomb potential 
$v_0(q)=(2\pi e^2)/\epsilon q$ in Eqs. 
(\ref{VH})-(\ref{VF}) replaced by 
$v(q) = (2\pi e^2)e^{-\lambda q\ell}/\varepsilon(q)q$.
The parameter $\lambda$ models a finite thickness sample,\cite{Zhang} and is generally taken to 
be of order unity. Here we use $\lambda=1$, while for 
$\varepsilon(q)$ we shall use the following expression, due to Aleiner and 
Glazman,\cite{Aleiner}
$
\varepsilon(q) =\epsilon\Big(
1 + \frac{2}{q a_B}[1-J_0^2(qR_c)]
\Big) \,,
$
with $a_B=\hbar^2\epsilon/m^*e^2$ the effective Bohr radius and $R_c=\sqrt{2n+1}\,\ell$.
We find that, although screening and finite sample thickness shift the cohesive energies up, the 
relative boundaries between the various phases remain practically unchanged.

\section{Conclusion}

In conclusion, in this paper we have argued that a new bubble crystal with a basis may be 
an energetically more favorable bubble ground state of the 2D electron system.
We have shown that this new bubble state with basis is stable against small fluctuations of the 
guiding center positions, and remains energetically favorable in finite thickness samples and if 
we take screening by lower LLs into account. More detailed investigations of other possible crystal 
structures for the Wigner crystal and bubble phase (along with experimental consequences) are 
under way and 
will be discussed elsewhere.

\section*{Acknowledgements}

This work was supported by the NHMFL In House Research Program.

\end{document}